**Power Spectrum Analysis of BNL Decay-Rate Data**

P.A. Sturrock[a,*], J.B. Buncher[b], E. Fischbach[b], J.T. Gruenwald[b], D. Javorsek II[c], J.H. Jenkins[b], R.H. Lee[d], J.J. Mattes[b], J.R. Newport[b]

[a] Center for Space Science and Astrophysics, Stanford University, Stanford, CA 94305-4060, USA

[b] Department of Physics, Purdue University, West Lafayette, IN 47907, USA

[c] 416[th] Flight Test Squadron, 412[th] Test Wing, Edwards AFB, CA 93524, USA

[d] Department of Physics, United States Air Force Academy, CO 80920, USA

- Corresponding author. Tel +1 6507231438; fax +1 6507234840.

Email address: sturrock@stanford.edu

**Abstract**

Evidence for an anomalous annual periodicity in certain nuclear decay data has led to speculation concerning a possible solar influence on nuclear processes. As a test of this hypothesis, we here search for evidence in decay data that might be indicative of a process involving solar rotation, focusing on data for $^{32}$Si and $^{36}$Cl decay rates acquired at the Brookhaven National Laboratory. Examination of the power spectrum over a range of frequencies (10 – 15 year$^{-1}$) appropriate for solar synodic rotation rates reveals several periodicities, the most prominent being one at 11.18 year$^{-1}$ with power 20.76. We evaluate the significance of this peak in terms of the false-alarm probability, by means of the shuffle test, and also by means of a new test (the "shake" test) that involves small random time displacements. The last two tests are the more robust, and indicate that the peak at 11.18 year$^{-1}$ would arise by chance only once out of about $10^7$ trials. However, the fact that there are several peaks in the rotational search band suggests that modulation of the count rate involves several low-Q oscillations rather than a single high-Q oscillation, possibly indicative of a partly stochastic process. To pursue this possibility, we investigate the running mean of the power spectrum, and identify a major peak at 11.93 year$^{-1}$ with peak running-mean power 4.08. Application of the shuffle test indicates that there is less than one chance in $10^{11}$ of finding by chance a value as large as 4.08. Application of the shake test leads to a more restrictive result that there is less than one chance in $10^{15}$ of finding by chance a value as large as 4.08. We find that there is notable agreement in the running-mean power spectra in the rotational search band formed from BNL data and from ACRIM total solar irradiance data. Since rotation rate estimates derived from irradiance data have been found to be closely related to rotation rate estimates derived from low-energy solar-neutrino data, this result supports the recent conjecture that solar neutrinos may be responsible for variations in nuclear decay rates. We also carry out a similar comparison with local temperature measurements, but find no similarity between power spectra formed from BNL measurements and from local temperature measurements.

## 1. Introduction

There is strong evidence [1-3] for an annual periodicity in decay data acquired at both the Brookhaven National Laboratory (BNL) [4] and at the Physicalisch Technische Bundesandstalt (PTB) [5]. Such a variation could in principle be caused by environmental effects, but a detailed investigation by our collaboration shows that the results of our BNL and PTB analyses cannot be explained by variations of temperature, pressure, humidity, etc. [6]. Norman et al. [7] have reexamined data from several studies of nuclear decay rates and found no evidence for a correlation with Sun-Earth distance. However, our collaboration has recently re-analyzed Norman's data, which Norman and his collaborators generously provided, and we have detected an annual periodicity with a small amplitude but the same phase as that found in the BNL and PTB datasets. Cooper [8] has analyzed data from the power output of the radioisotope thermoelectric generators aboard the Cassini spacecraft, finding no significant deviations from exponential decay. However, we show in Section 9 of our recent article [6] that there is no conflict between Cooper's results and our results [1-3].

To investigate more closely the possibility that the annual variations are somehow caused by the Sun, we now look for a periodicity that may be associated with solar rotation. It is convenient to adopt a working hypothesis for a possible solar influence. If the data prove to be incompatible with the hypothesis, that will reduce by one the number of possible causes. If the data were to prove to be compatible with the hypothesis, that would strengthen the case for that hypothesis. We adopt, as a working hypothesis, our recent suggestion that nuclear decay rates are influenced by solar neutrinos [1].

The solar convection zone and radiative zone have synodic rotation frequencies (as observed from Earth) in the equatorial plane in the range 12.9 to 13.8 year$^{-1}$ (corresponding to sidereal rotation frequencies 13.9 to 14.8 year$^{-1}$) [9]. However, we have found evidence that low-energy solar neutrinos, as detected by the Homestake [10,11] and GALLEX [12-15] experiments, are modulated at a lower frequency (11.85 year$^{-1}$) that shows up also in ACRIM [16,17] total solar irradiance data for the Homestake and GALLEX time intervals [18,19]. It is reasonable to suspect that this frequency may be related to the synodic rotation frequency of the solar core, since this is where neutrinos originate. These results suggest that the solar core rotates more slowly than either the convection zone or the radiative zone, and that rotation of the core influences both the solar neutrino flux and (somewhat surprisingly) the solar irradiance.

The primary goal of this article is therefore to determine whether decay data—specifically the BNL data—show evidence of a periodicity in a "search band" appropriate for synodic solar rotation rates, which we take to be 10—15 year$^{-1}$. For the above reasons, we shall be particularly interested in evidence for rotation rates that are slower than those of the radiative zone and the convection zone.

The preparation of the data and the power spectrum analysis are presented in Section 2, where we find a major peak at frequency 11.18 year$^{-1}$, and assess the significance of that peak by calculating the false-alarm probability [20] and (in Section 3) by applying the more robust shuffle test [21]. We also carry out a significance estimate using a new test that we call the "shake test." This involves the comparison of the actual power spectrum with a large number of power spectra derived from simulated data generated by small random displacements of the times of measurements.

However, the power spectrum formed in Section 2 contains several peaks in the rotational search band, suggesting that the variations in the decay rates are not to be attributed to a unique high-Q (stable, narrow-band) oscillation (due for instance to the steady rotation of a steady but asymmetric source). It may be due to several transient oscillations of different frequencies, or to an oscillation that drifts in amplitude and frequency, or to a process that is partly periodic and partly stochastic. For this reason, we seek to evaluate the entire power spectrum over the rotational search band. To this end, we form in Section 4 a weighted running mean of the power spectrum. We find a major peak at 11.93 year$^{-1}$, very close to the major peak (at 11.85 year$^{-1}$) found in a comparative analysis of low-energy solar neutrino data and total solar irradiance data [18,19]. We examine the significance of this peak in Section 5, using the shuffle and shake tests. These tests indicate that the power-spectrum structure found in the rotational search band is highly significant.

We discuss these results in Section 5, where we compare the running-mean power spectrum formed from BNL data with those formed from ACRIM irradiance data [16,17] and from local temperature measurements. We find a close similarity with the former, but no similarity with the latter.

## 2. Power Spectrum Analysis

The BNL dataset comprises 366 measurements in the time interval 1982.11 to 1989.93, where dates are measured in what we call "neutrino years," which have proved useful in analyzing neutrino data. (Dates in "neutrino days" are counted from January 1, 1970, as day 1, and dates in "neutrino years" are given by 1970 + (neutrino days)/365.2564.) We have prepared the data for time-series analysis by dividing the decay count rates by the counts expected on the basis of a purely exponential decay, using the mean decay rate determined from the best fit to the data. The resulting normalized count rates are shown in Figures 1 and 2 for the Cl ($^{36}$Cl) and Si ($^{32}$Si) data, respectively. We show in Figure 3 the ratio of the normalized Si measurements to the normalized Cl measurements. It is clear that the ratio of Si to Cl measurements shows less scatter than either the Cl or the Si measurements. This is borne out by inspection of the standard deviations, which are 0.0013 and 0.0014 for Cl and Si data, respectively, but only 0.0011 for the ratio. This comparison suggests that the Cl and Si measurements are subject to some experimental (environmental or instrumental) influences which are mitigated on forming the ratio. For the above reasons, we have focused our attention on the ratio of the measurements.

We have next performed a power-spectrum analysis of the data shown in Figure 3, using a likelihood procedure [22] that is equivalent to the Lomb-Scargle procedure [20, 23]. The result is shown in Figure 4. The biggest peak is at 0.17 year$^{-1}$, which should probably be interpreted as a secular trend, since the dataset extends over less than 8 years. The next biggest peak is close to 1 year$^{-1}$, which is to be expected [1-3]. The third, fourth, and fifth peaks (11.17 year$^{-1}$ with power S = 20.76; 13.11 year$^{-1}$ with S = 16.79; and 11.73 year$^{-1}$ with S = 12.96) all fall within the band of 10—15 year$^{-1}$, which we adopt as our search band for rotational frequencies.

There are 22 peaks in the search band. Adopting this value for M, the number of "independent measurements" in that band, and using the standard formula [20] for the "false-alarm probability",

$$FAP = 1 - \left(1 - e^{-S}\right)^{M},$$  (1)

with S = 20.76, we obtain the estimate FAP = 2.1 10$^{-8}$. However, when one is over-sampling a power spectrum (as in this case), the average value of the peaks is larger than the average value of the power by approximately unity [24]. Using $S_M$ = 19.76 and N = 22, we obtain the revised estimate FAP = 5.8 10$^{-8}$.

### 3. Significance Estimates

The false-alarm formula is derived on the assumption that measurements are derived from an exponential distribution, as would be appropriate for a power spectrum derived from normally distributed random noise [18]. This may be approximately true for the current dataset, but there is no reason to expect that it is absolutely true. Since the false-alarm probability estimate depends critically on this assumption, it is advisable to perform more robust significance estimates that do not depend on that assumption. One such procedure is the "shuffle test" [21].

We denote by $t_n$, and $x_n$, n = 1,…,N, the times of the measurements and the measurements, respectively. In carrying out the shuffle procedure, we retain the actual values $t_n$ and $x_n$, but we pair up the times and the measurements randomly. This test does not depend on an assumed form for the distribution of measurements. For each simulation, we repeat the power spectrum analysis, and note the value of the biggest peak in the search band. The result of 10,000 such simulations is shown in histogram form in

Figure 5. For none of the 10,000 simulations is there a peak in the search band as large as the actual power (20.76). (The largest value is found to be 14.16.)

These results are also shown in a logarithmic display in Figure 6. This figure also shows an extension of the empirical curve, derived from a projection of the last 1,000 points. This projection leads us to expect that we would obtain a peak as big as the actual peak only once in about $10^7$ random simulations.

We have already noted that there are two peaks in the power spectrum shown in Figure 4 that are bigger than the peaks in the search band. The one at 0.17 year$^{-1}$ probably represents a secular trend. Since the shuffle test necessarily pairs measurements from times that are well removed from each other, this could conceivably introduce a bias in the significance estimate derived from the shuffle test. We have therefore looked into the possibility of devising a test that is in principle similar to the shuffle test, but which would take into account the fact that the distribution of measurements may not be the same throughout the dataset. We believe that the following "shake test" meets these requirements.

We retain the actual values $x_n$ of the measurements. However, we replace each time $t_n$ by one that differs from $t_n$ by an amount that is small compared with the timescale for changes in the pattern of measurements, but is big enough to disrupt periodicity in the range being investigated. Specifically, we use the algorithm

$$t_n \rightarrow t_n + Dt * R \ , \qquad\qquad (2)$$

where $R$ denotes a number drawn randomly over the range -1 to 1. To decide on the appropriate value of R, we have run 1,000 analyses of a fictitious dataset comprising the actual periodicity at 11.18 year$^{-1}$ superposed on Gaussian noise with the same standard deviation as the actual data. We find that Dt = 0.0005 gives a power distribution close to a delta function at S = 20; Dt = 0.005 gives a wide scatter from S = 3 to S = 21; Dt = 0.05 gives a distribution close to that obtained from S = 0.5, a bell-shaped distribution ranging from S = 3 to S = 10. We have therefore adopted the value Dt = 0.05 year, which is large enough to disrupt the coherence of a wave with frequency in the range 10 - 15 year$^{-1}$, but small compared with the timescale (of order 1 year) of the patterns seen in Figure 3.

We have carried out 10,000 shake simulations of the data, arriving at the result shown in histogram form in Figure 7. One again, we find that none of the 10,000 simulations leads to a power as large as the actual power (20.76) of the principal peak in the search band. (The largest value is 12.92.)

These results are shown in a logarithmic display in Figure 8, which also shows an extension of the empirical curve, derived from a projection of the last 1,000 points. This projection leads us to expect that we would obtain a peak as large or larger than the actual peak only once in about $10^{7.4}$ simulations. This would correspond to a false-alarm probability FAP = 4.0 $10^{-8}$, close to the estimate 5.8 $10^{-8}$ obtained in Section 2.

## 4. Frequency-Averaged Power Spectra

We see from the power spectrum shown in Figure 4 that there are several peaks in the range 11 - 13 year$^{-1}$, indicating that the time series should not be attributed to a single stationary high-Q periodic process. It suggests, if we think in terms of variability due primarily to rotation, that the signal may be attributed to rotation in more than one region, and that the rotation rate of any region may be variable. (It is also worth noting that the power spectrum produced by a hypothetical obliquely rotating Sun depends critically on the angle between the rotation axis and the normal to the ecliptic [25].) For this reason, we seek a procedure for evaluating a complex of peaks, rather than focus on only one peak. To this end, we have adopted the procedure of forming a running mean of the power spectrum.

We denote by $\nu_j$ the sequence of frequencies (with spacing 0.01 year$^{-1}$) at which the power is evaluated, and by S$_j$ the power sequence, where $j = 1,...,N_\nu$. We then form the sequence

$$\tilde{S}_j = \sum_k W(j-k) S_k \ . \tag{3}$$

We have adopted

$$W(j) = C \cos\left(\frac{\pi}{2} \frac{j}{m}\right), \quad j = -m, ..., m \ , \tag{4}$$

in which C is chosen so that the mean value of W is unity:

$$C = \sum_{-m}^{m} W(j) \ . \tag{5}$$

We have adopted m = 200, which is equivalent to forming running means of the power over intervals of width 4 year$^{-1}$. However, we find that, for m = 200,

$$\left(\sum_{-m}^{m} W(j) j^2\right)^{1/2} = 0.435 \, m \ , \tag{6}$$

so that the rms frequency deviation is only 0.87 year$^{-1}$.

The result of applying this smoothing operation to the power spectrum shown in Figure 4 (extended to 100 year$^{-1}$) is shown in Figure 9. We find that the smoothed power spectrum has a peak centered on 11.93 year$^{-1}$ with peak power 4.08.

In order to obtain a significance estimate for this peak, we again use the shuffle test, re-computing the power spectrum and then the running mean of the power spectrum many times, keeping the original times and the original count rates, but randomly sorting one of these datasets. We have carried out this procedure 10,000 times, determining the maximum weighted running mean of the power in the rotational search band for each simulation. The histogram formed from these maxima is shown in Figure 10. We see that none of the random simulations yields as large a value of the running mean of the power in the search band as the actual value (4.08). (The largest value is 2.11.)

These results are shown in a logarithmic display in Figure 11, which also shows an extension of the empirical curve, derived from a projection of the last 1,000 points. This projection leads us to a p-value of $10^{-11.0}$, implying that we could expect to obtain a peak as large or larger than the actual peak only once in about $10^{11}$ simulations.

We have also examined the running-mean power spectrum by means of the shake test, again carrying out 10,000 simulations. The results are shown in histogram form in Figure 12 and as a logarithmic display in Figure 13. We see that none of the simulations has as large a value of the maximum running-mean power in the search band as we obtain in our analysis of the actual data. (The largest value is 2.02.) A projection of the curve is found to have the abscissa value 4.08 where the ordinate value is – 15.5, from which we infer that the result of this test corresponds to a p-value of $10^{-15.5}$, a stricter result than that found from the shuffle test.

## 5. Discussion

We have recently drawn attention to the fact that a p-value may not be interpreted as the probability that the null hypothesis is false, and we have proposed a Bayesian approach that does lead to such a probability [26]. Equation (25) of that article leads to the following formula for the odds on the null hypothesis,

$$\Omega\left(H_0 \mid pv\right) = 2.44\left(1.92 - \log\left(pv\right)\right)pv\,. \qquad (7)$$

If we adopt the more conservative estimate pv $= 10^{-11.0}$ from the shuffle test in Section 4, rather than the value derived from the shake test, we estimate the odds on the null hypothesis to be 7 $10^{-10}$, implying that the odds are 1.4 $10^9$ in favor of a real modulation in the rotational search band.

In the Introduction, we drew attention to the fact that a combined analysis of low-energy solar neutrino data and the total solar irradiance shows that neutrinos and irradiance exhibit a periodicity at the same frequency, 11.85 year$^{-1}$[18,19]. However, inspection of the power spectra derived from the neutrino and irradiance data shows that this is simply the dominant peak of a cluster of peaks. Hence, in order to compare the BNL data with neutrino and irradiance data, it is convenient to analyze either neutrino or irradiance data in the same way that we have analyzed decay data, i.e. in terms of running-mean power-spectra. It is more advantageous to work with irradiance data since they are more accurate and extensive, so that we may extract measurements for the BNL time interval.

Beginning with ACRIM measurements [16,17] over the time interval of BNL measurements, we have carried out a power spectrum analysis and then formed the weighted running means of the power using the procedure outlined in Section 4. The result of this analysis is shown in Figure 14 in which we also reproduce for comparison the curve derived from BNL data, previously shown in Figure 9. Focusing on the rotational search band 10 – 15 year$^{-1}$, we see that these distinct datasets both yield significant peaks at essentially the same frequency, approximately 11.9 year$^{-1}$.

As we noted in the introduction, such a low frequency cannot be attributed to either the radiative zone or the convection zone. This leaves only the solar core as the place of origin of this periodicity. This of course helps one understand the association with solar neutrinos, since the core is by definition the location of nuclear reactions in the Sun.

(We remark, parenthetically, that the peak in the irradiance curve at 27.2 year$^{-1}$ is too high to be a harmonic of the periodicity associated with the core, but it may be interpreted as a harmonic of the photospheric rotation frequency for which the equatorial synodic frequency is in the range 13.4—13.6 year$^{-1}$, suggesting that it may be attributed to an ellipsoidal distortion of the photosphere.)

In reviewing evidence for an annual oscillation in decay data [1-3], one faces the requirement of distinguishing a periodicity in the data due to nuclear processes from possible periodicities due to environmental influences. Of the possible environmental factors (temperature, pressure, humidity, etc.), the local ambient temperature shows the strongest seasonal variation. We have therefore applied our power-spectrum running-mean analysis to the local temperature associated with the BNL experiment. The result is shown in Figure 15, where we again reproduce for comparison the curve for BNL data shown in Figure 9. We see that (in contrast to our results from decay data and irradiance data) the ambient temperature does not lead to a corresponding peak in the rotational search band. It therefore seems highly unlikely that environmental effects have played any role in leading to a periodicity in the rotational search band.

We have found evidence that decay data exhibit a periodicity very close to that found in low-energy solar neutrino data and in solar irradiance data. Hence our results are compatible with our "working hypothesis" for a solar influence on nuclear decay rates. Concerning possible mechanisms, it has seemed reasonable to attribute periodicities in neutrino and irradiance data in the range 11 – 12 year$^{-1}$ to processes involving the rotation of a longitudinally asymmetric solar core [18,19]. It is worth noting that, since the Sun's axis is inclined with respect to the normal to the ecliptic, a latitudinal asymmetry in the solar core could contribute to the annual variation found in decay rates [1-3]. This effect, in association with an annual variation due to the eccentricity of the Earth's orbit, may explain why the phases of the annual variations in decay data do not agree with what would be expected on the basis of a purely orbital effect. This issue will be discussed in more detail in a later article.

Although the physical implications of the present analysis remain uncertain, the results appear to be compatible with the hypothesis that some nuclear decay rates are influenced by solar neutrinos. We may not conclude from this compatibility that the hypothesis is valid, but we may conclude that—as far as the present analysis is concerned—the hypothesis remains viable.

## Acknowledgements

The authors are indebted to David Alburger and Garman Harbottle for supplying us with raw data from the BNL experiment; to Claus Frohlich and Judith Lean for providing solar irradiance data; and to Jeffrey D. Scargle and Guenther Walther for helpful advice on statistical issues. The work of PAS was supported in part by the National Science Foundation through grant AST-0097128 . The views expressed in this paper are those of the authors and do not reflect the official policy or position of the U.S. Air Force, the U.S. Department of Defense, or the U.S. Government.

# FIGURE CAPTIONS

Figure 1. Normalized count rates from the $^{36}$Cl source in the BNL experiment.

Figure 2. Normalized count rates from the $^{32}$Si source in the BNL experiment.

Figure 3. Ratio of the normalized count rate from the $^{32}$Si source to the normalized count rate from the $^{36}$Cl source in the BNL experiment.

Figure 4. Power spectrum formed by a likelihood procedure from the ratio of the Si and Cl count rates.

Figure 5. Histogram of 10,000 power estimates generated by the shuffle procedure. None has power as large as that in the actual power spectrum (20.76, shown in red; the largest value is 14.16).

Figure 6. Logarithmic display of the power estimates generated by 10,000 shuffle simulations of the BNL data. A projection of the curve indicates that one would expect to obtain the actual power (20.76) only once in about $10^7$ trials.

Figure 7. Histogram of 10,000 power estimates generated by the shake procedure. None has power as big as that in the actual power spectrum (20.76, shown in red; the biggest value is 12.92).

Figure 8. Logarithmic display of the power estimates generated by 10,000 shake simulations of the data. A projection of the curve indicates that one would expect to obtain the actual power (20.76) once in about $10^7$ trials.

Figure 9. Plot of the 501-point weighted running means formed from the power spectrum shown in Figure 4 . The peak occurs at frequency 11.93 year$^{-1}$ with weighted-running-mean power 4.08.

Figure 10. Histogram of 10,000 estimates of the maximum weighted-running-mean power in the search band 8—17 year$^{-1}$ generated by the shuffle procedure. None of the simulations gives a value as big as that (4.08) in the actual data. (The biggest value is 2.11.)

Figure 11. Logarithmic display of 10,000 estimates of the maximum weighted-running-means power in the search band 8—17 year$^{-1}$ generated by the shuffle procedure. A projection of the curve indicates that one would expect to obtain the actual value (4.08) only once in about $10^{11}$ trials.

Figure 12. Histogram of 10,000 maximum weighted-running-mean power generated by the shake procedure. None has a value as big as that derived from the actual data (4.08). (The biggest value is 2.02.)

Figure 13. Logarithmic display of the maximum weighted-running-mean power generated by 10,000 shake simulations of the BNL data. A projection of the curve indicates that one would expect to obtain the actual value (4.08) less than once in about $10^{15}$ trials.

Figure 14. Plot of the weighted-running-mean power formed from the BNL data (red), and the corresponding figure formed from the ACRIM irradiance data for the BNL time interval (green). The ACRIM power has been reduced by a factor of 2.

Figure 15. Plot of the weighted-running-mean power formed from the BNL data (red), and the corresponding figure formed from local temperature data for the BNL time interval (green).

FIGURES

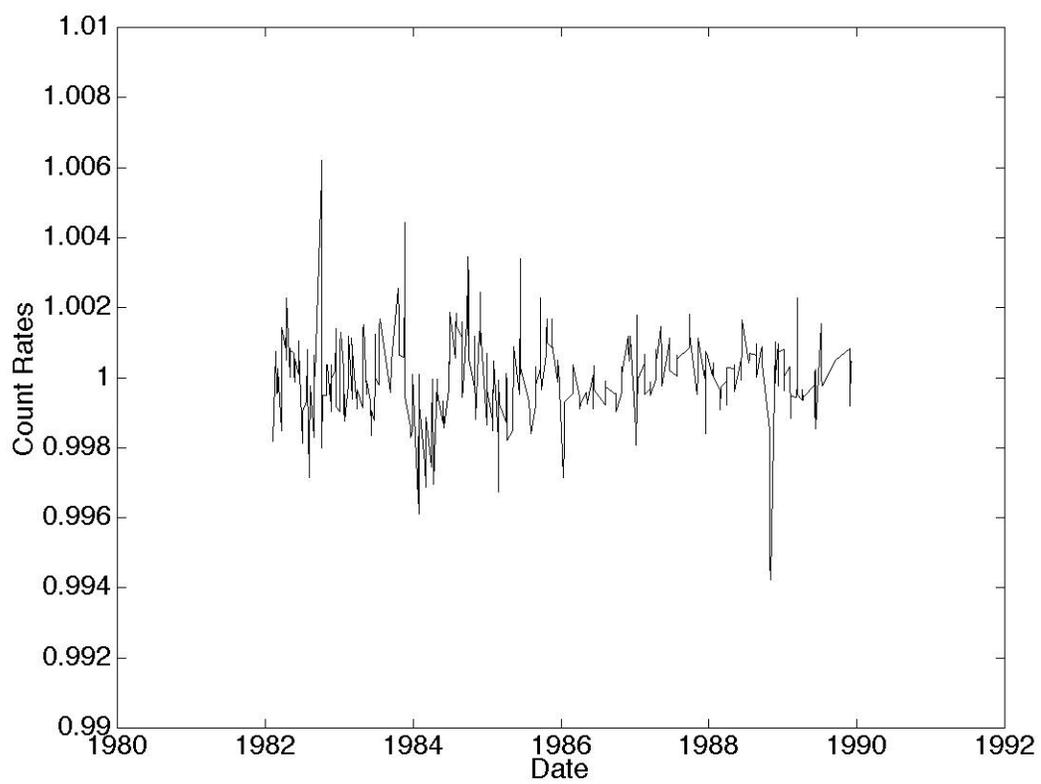

Figure 1. Normalized count rates from the $^{36}$Cl source in the BNL experiment.

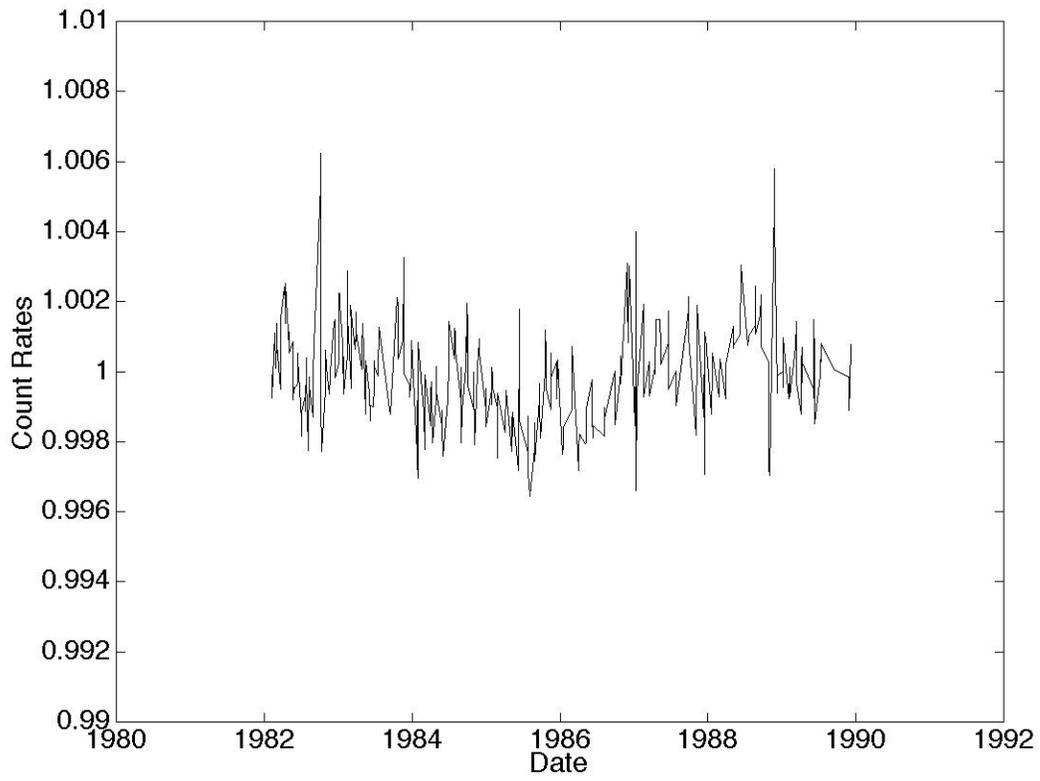

Figure 2. Normalized count rates from the $^{32}$Si source in the BNL experiment.

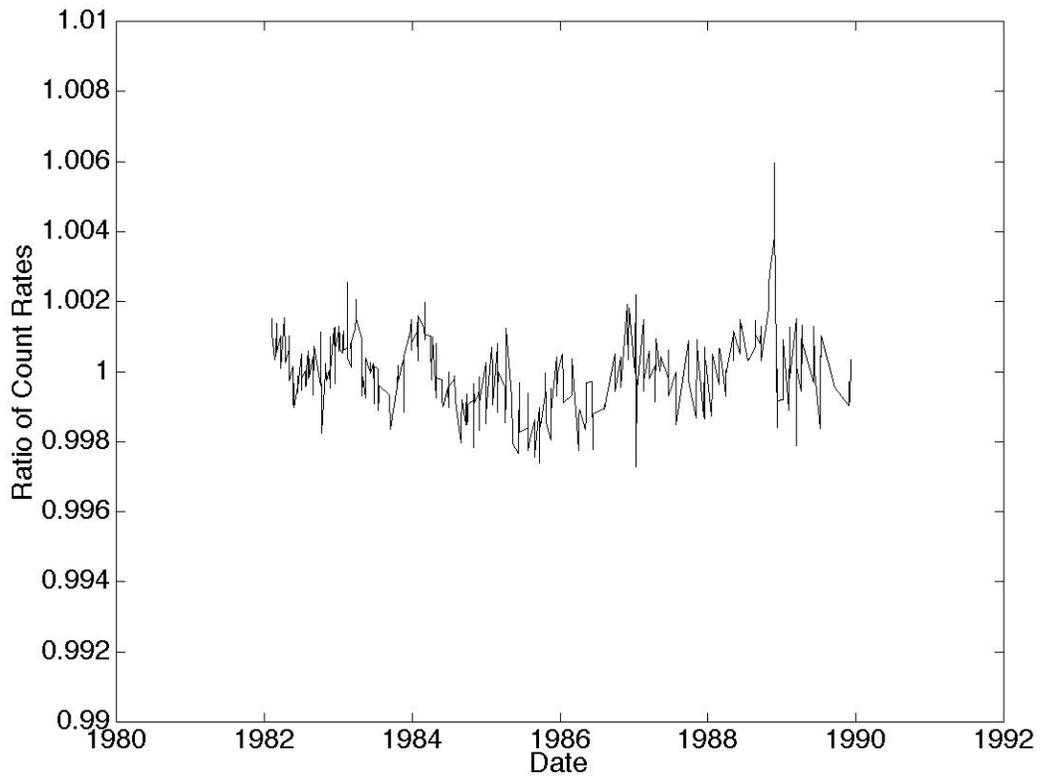

Figure 3. Ratio of the normalized count rate from the $^{32}$Si source to the normalized count rate from the $^{36}$Cl source in the BNL experiment.

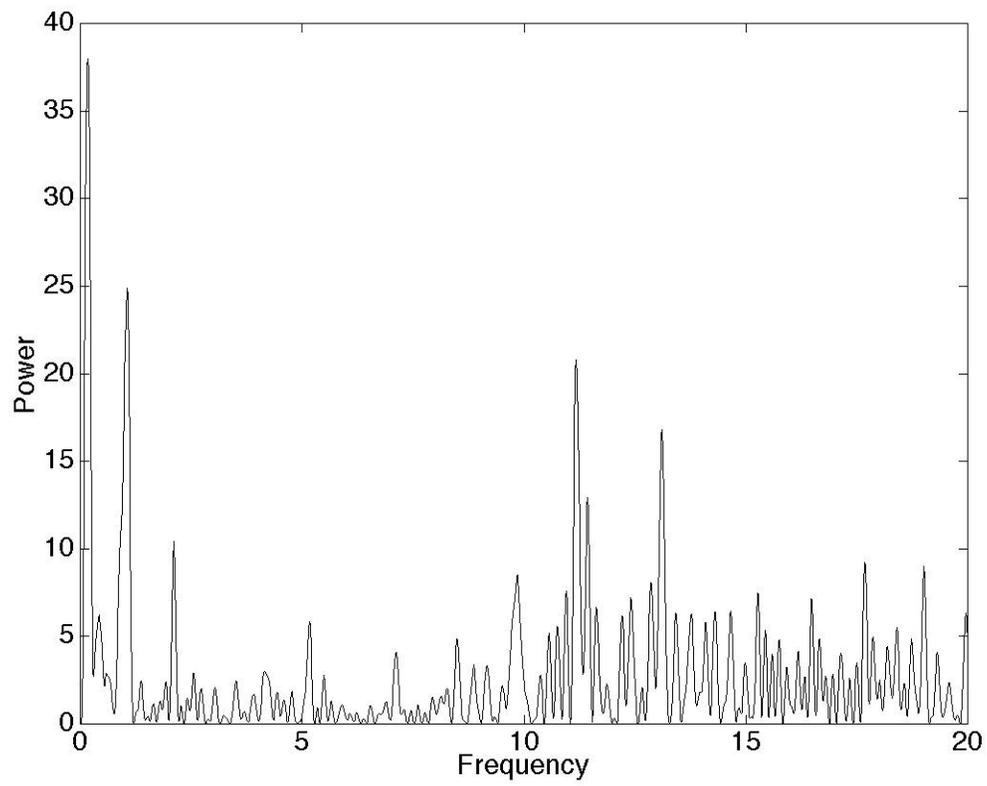

Figure 4. Power spectrum formed by a likelihood procedure from the ratio of the Si and Cl count rates.

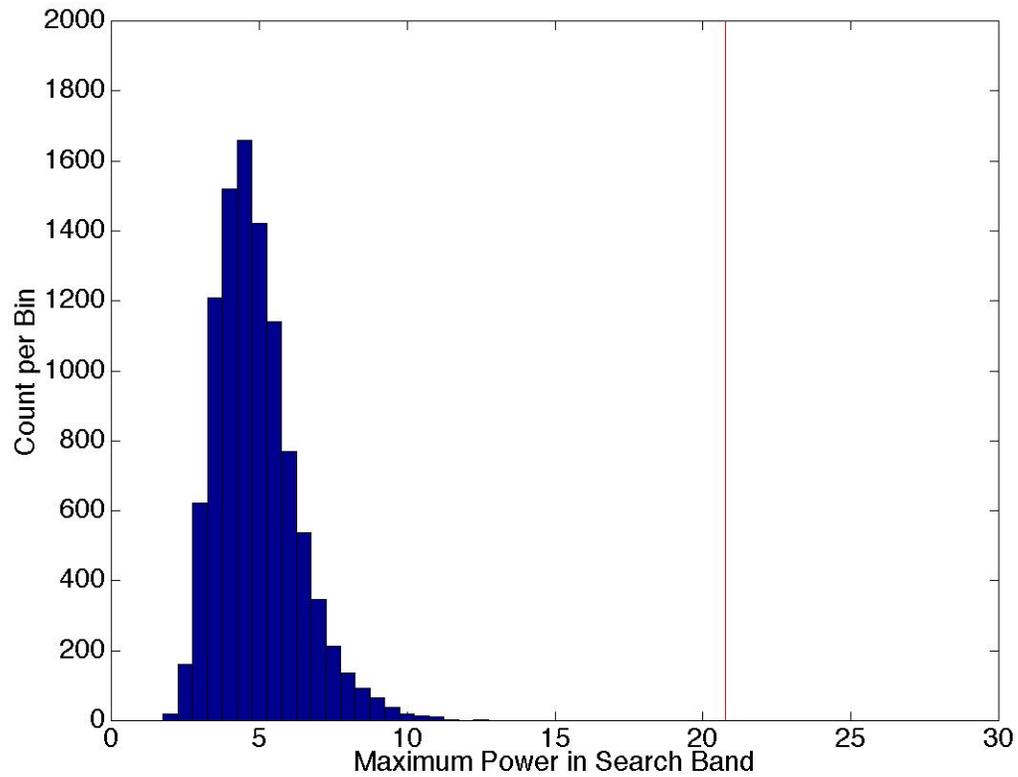

Figure 5. Histogram of 10,000 power estimates generated by the shuffle procedure. None has power as large as that in the actual power spectrum (20.76, shown in red; the largest value is 14.16).

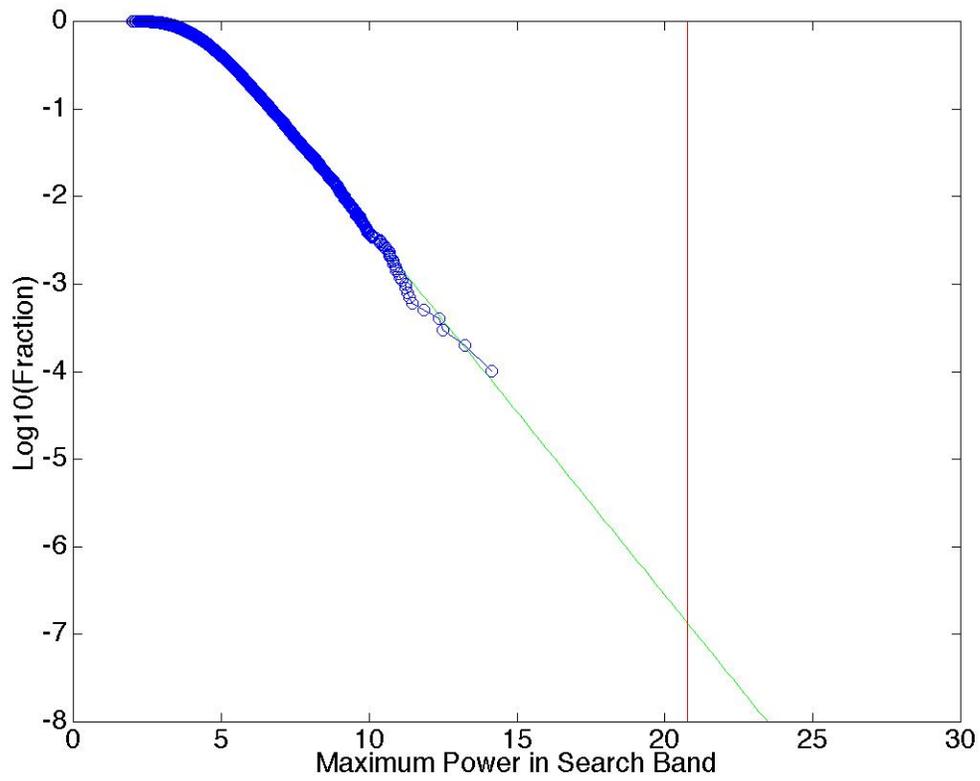

Figure 6. Logarithmic display of the power estimates generated by 10,000 shuffle simulations of the BNL data. A projection of the curve indicates that one would expect to obtain the actual power (20.76) only once in about $10^7$ trials.

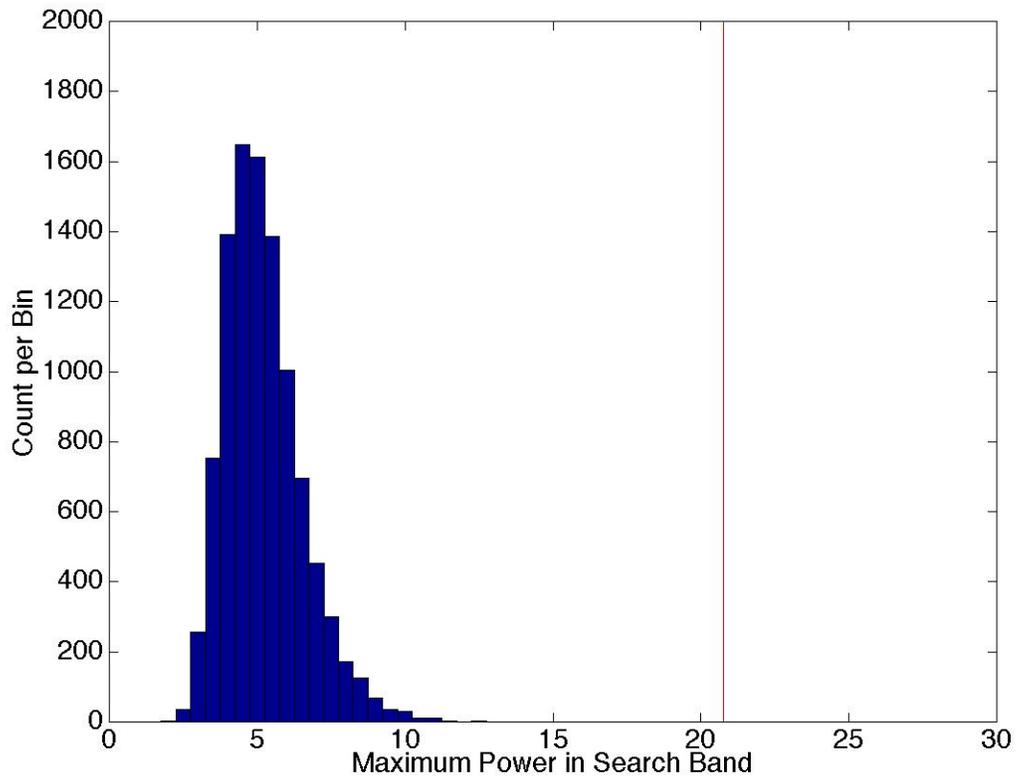

Figure 7. Histogram of 10,000 power estimates generated by the shake procedure. None has power as big as that in the actual power spectrum (20.76, shown in red; the biggest value is 12.92).

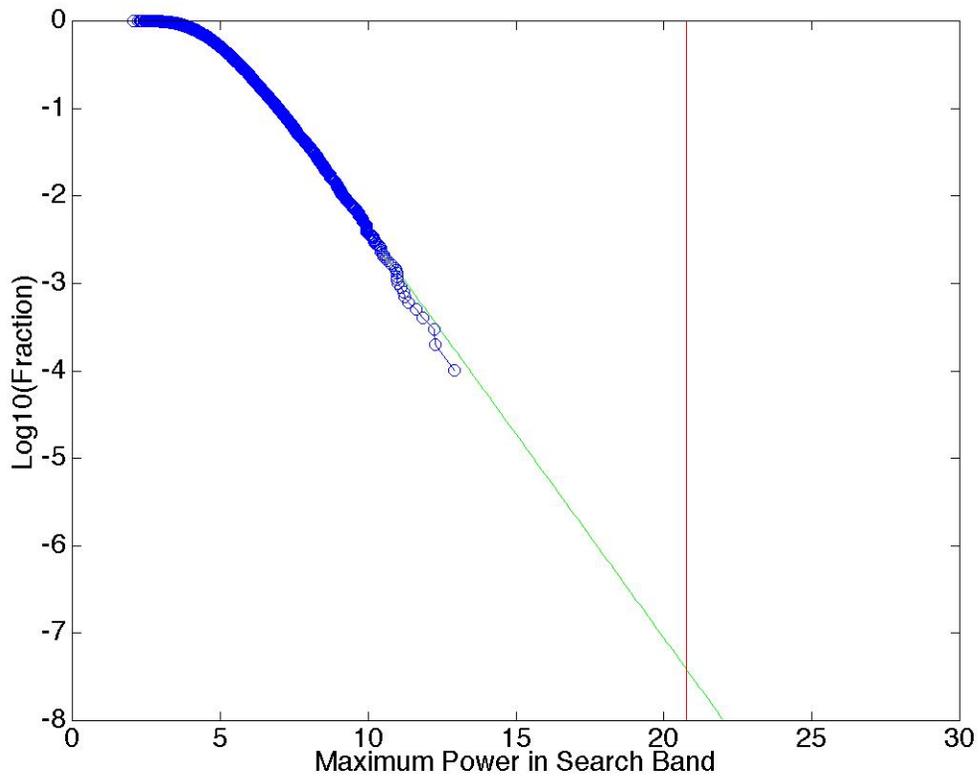

Figure 8. Logarithmic display of the power estimates generated by 10,000 shake simulations of the data. A projection of the curve indicates that one would expect to obtain the actual power (20.76) once in about $10^7$ trials.

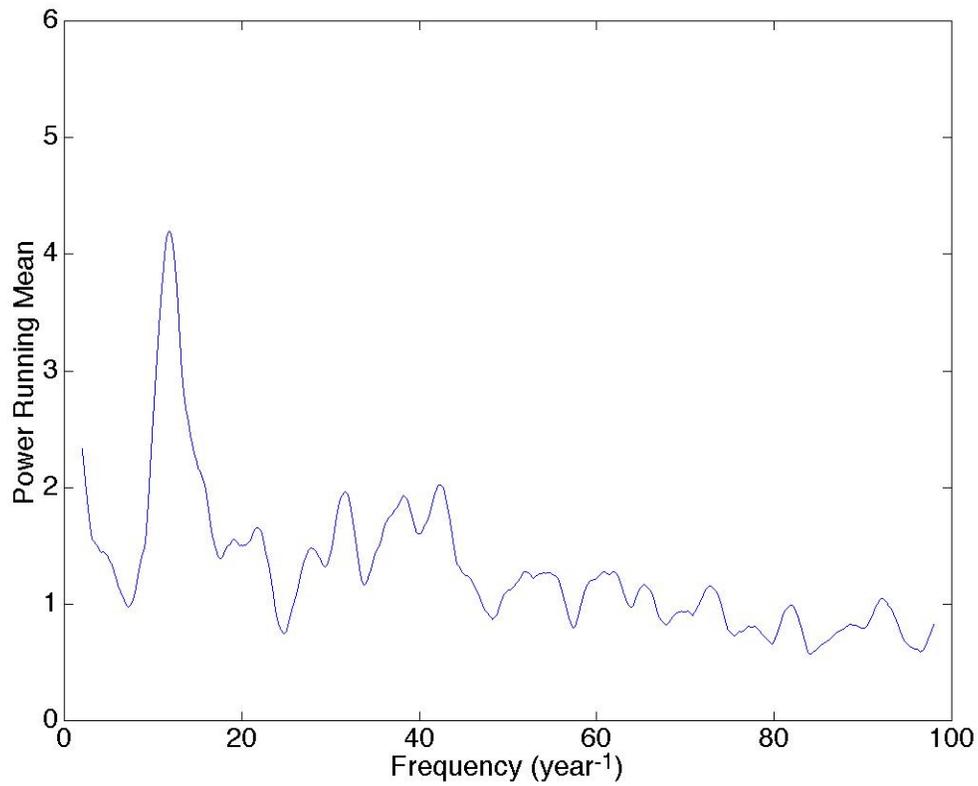

Figure 9. Plot of the 501-point weighted running means formed from the power spectrum shown in Figure 4 . The peak occurs at frequency 11.93 year$^{-1}$ with-weighted running-mean power 4.08.

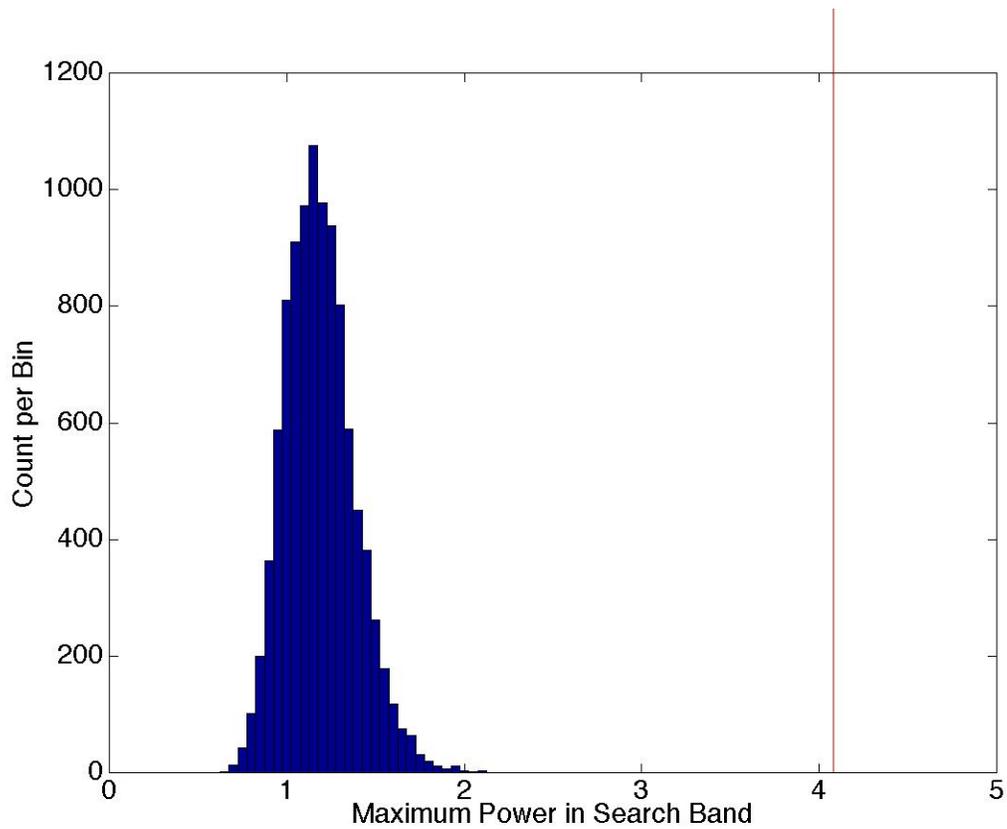

Figure 10. Histogram of 10,000 estimates of the maximum weighted-running-mean power in the search band 8—17 year$^{-1}$ generated by the shuffle procedure. None of the simulations gives a value as big as that (4.08) in the actual data. (The biggest value is 2.11.)

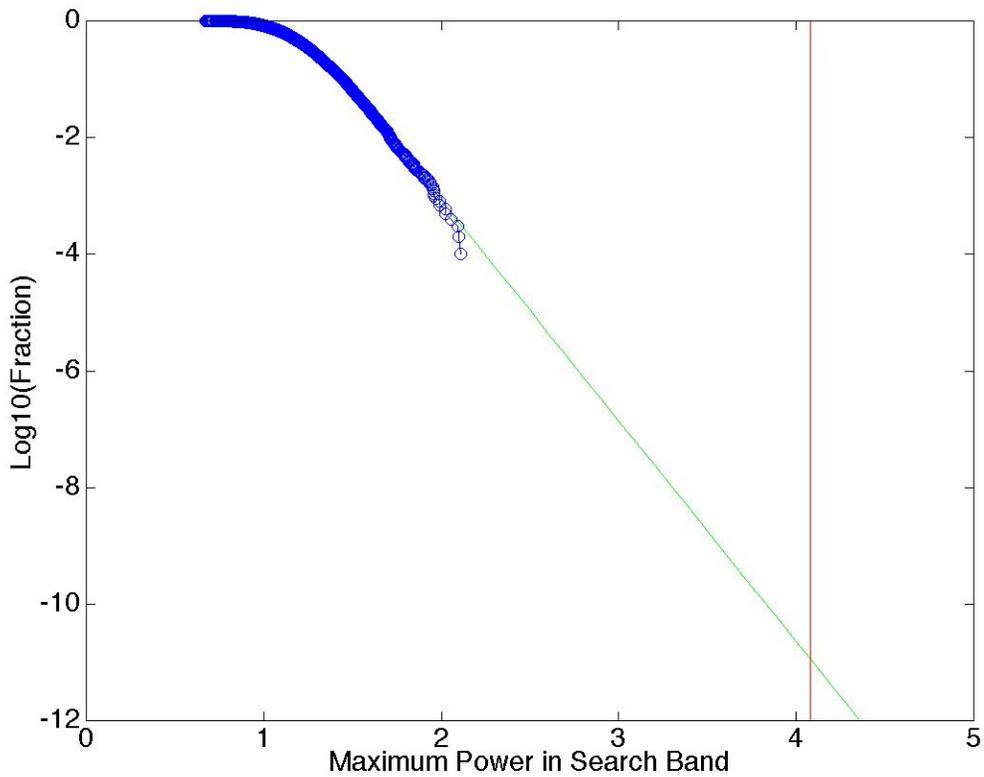

Figure 11. Logarithmic display of 10,000 estimates of the maximum weighted-running-means power in the search band 8—17 year$^{-1}$ generated by the shuffle procedure. A projection of the curve indicates that one would expect to obtain the actual value (4.08) only once in about $10^{11}$ trials.

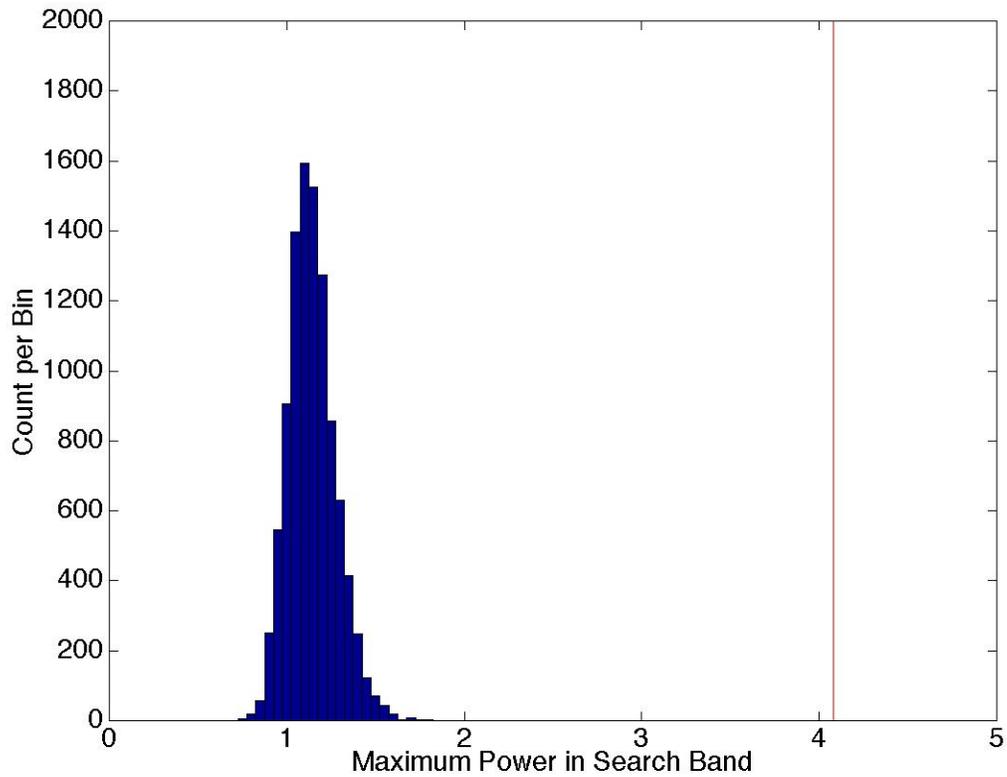

Figure 12. Histogram of 10,000 maximum weighted-running-mean power generated by the shake procedure. None has a value as big as that derived from the actual data (4.08). (The biggest value is 2.02.)

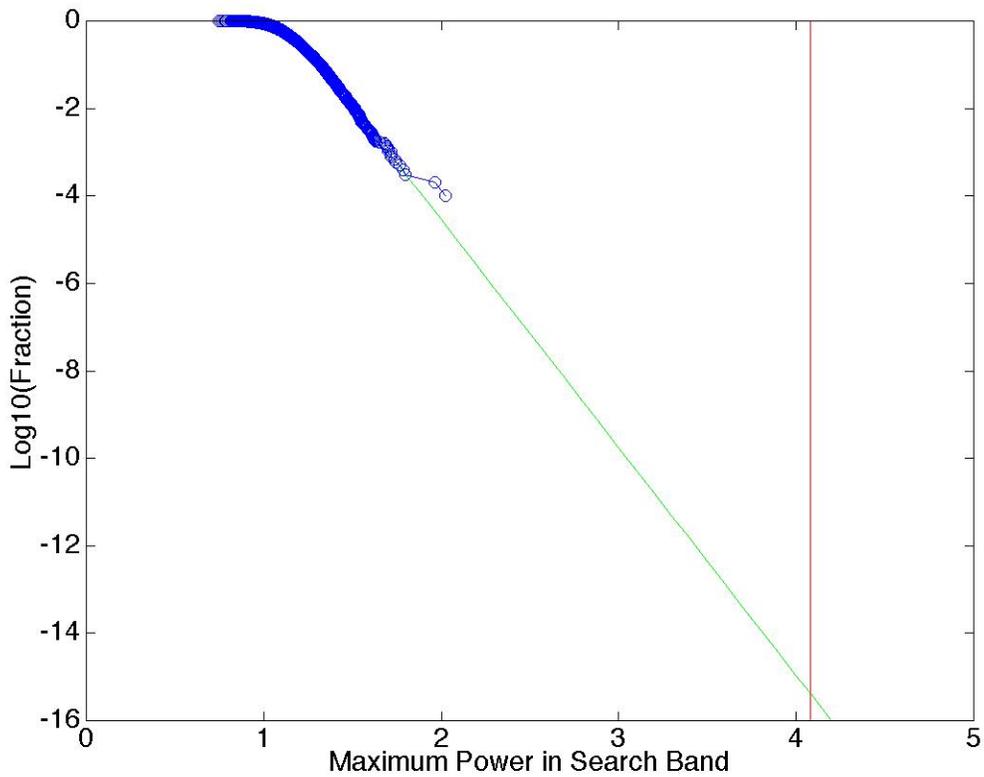

Figure 13. Logarithmic display of the maximum weighted-running-mean power generated by 10,000 shake simulations of the BNL data. A projection of the curve indicates that one would expect to obtain the actual value (4.08) less than once in about $10^{15}$ trials.

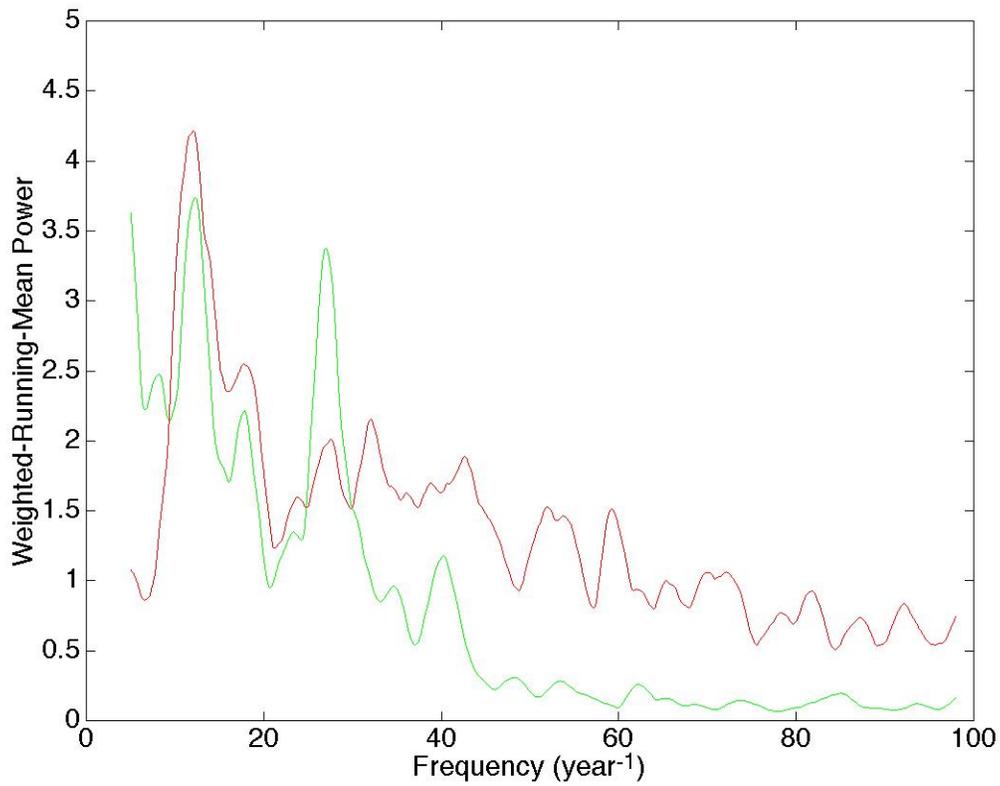

Figure 14. Plot of the weighted-running-mean power formed from the BNL data (red), and the corresponding figure formed from the ACRIM irradiance data for the BNL time interval (green). The ACRIM power has been reduced by a factor of 2.

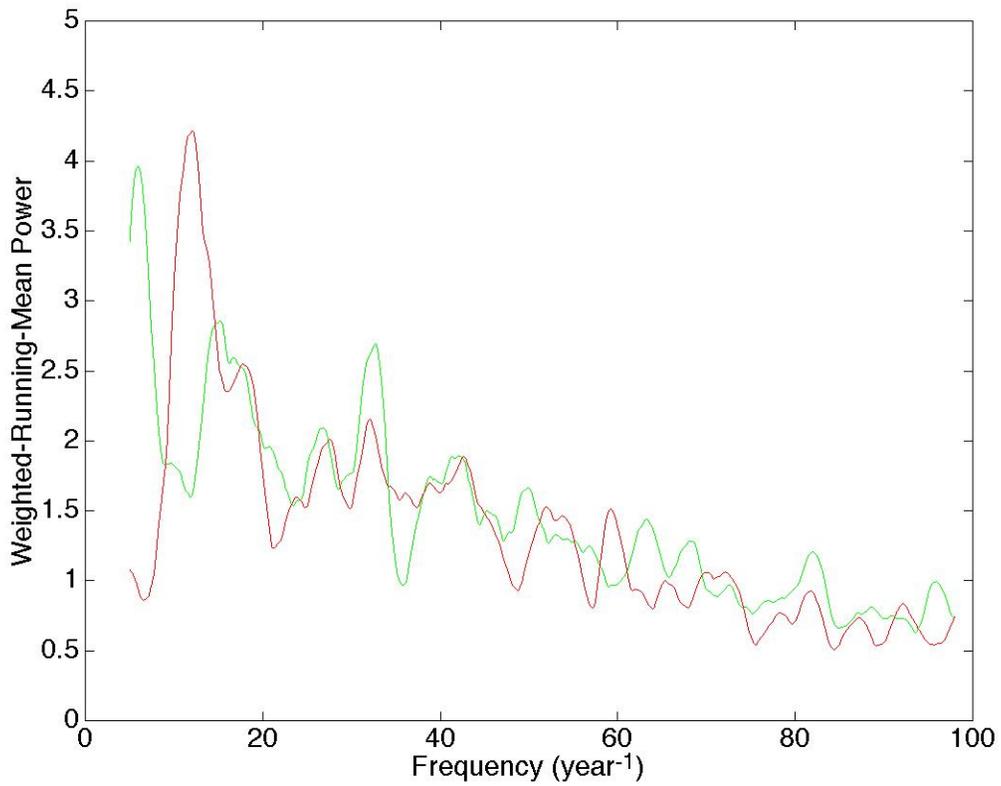

Figure 15. Plot of the weighted-running-mean power formed from the BNL data (red), and the corresponding figure formed from local temperature data for the BNL time interval (green).